\documentclass[twocolumn,aps,prl,superscriptaddress,preprintnumbers,amsmath,amssymb,nofootinbib]{revtex4-1}
\usepackage[T1]{fontenc}
\usepackage[latin9]{inputenc}
\usepackage{verbatim}
\usepackage{amssymb}
\usepackage{graphicx}
\usepackage[unicode=true,pdfusetitle,
 bookmarks=true,bookmarksnumbered=false,bookmarksopen=false,
 breaklinks=false,pdfborder={0 0 1},backref=false,colorlinks=false]
 {hyperref}
\makeatletter

 \@ifundefined{textcolor}{}
 {%
   \definecolor{BLACK}{gray}{0}
   \definecolor{WHITE}{gray}{1}
   \definecolor{RED}{rgb}{1,0,0}
   \definecolor{GREEN}{rgb}{0,1,0}
   \definecolor{BLUE}{rgb}{0,0,1}
   \definecolor{CYAN}{cmyk}{1,0,0,0}
   \definecolor{MAGENTA}{cmyk}{0,1,0,0}
   \definecolor{YELLOW}{cmyk}{0,0,1,0}
 }

\usepackage{color}
\usepackage{braket}

\def\0#1#2{\frac{#1}{#2}}

\def\s0#1#2{\mbox{\small{$ \frac{#1}{#2} $}}}

\newcommand{\be}{\begin{eqnarray}}
\newcommand{\ee}{\end{eqnarray}}

\makeatother

\begin{document}

\title{Atomic ionization of germanium due to neutrino magnetic moments}

\author{Jiunn-Wei Chen}
\affiliation{Department of Physics, National Taiwan University, Taipei 10617, Taiwan}
\affiliation{National Center for Theoretical Sciences and Leung Center for Cosmology
and Particle Astrophysics, National Taiwan University, Taipei 10617,
Taiwan}

\author{Hsin-Chang Chi}
\affiliation{Department of Physics, National Dong Hwa University, Shoufeng, Hualien
97401, Taiwan}

\author{Keh-Ning Huang}
\affiliation{Department of Physics, Sichuan University, Chengdu, Sichuan, China}
\affiliation{Department of Physics, Fuzhou University, Fuzhou, Fujian, China}
\affiliation{Department of Physics, National Taiwan University, Taipei 10617, Taiwan}

\author{C.-P. Liu}
\affiliation{Department of Physics, National Dong Hwa University, Shoufeng, Hualien
97401, Taiwan}

\author{Hao-Tse Shiao}
\affiliation{Department of Physics, National Taiwan University, Taipei 10617, Taiwan}

\author{Lakhwinder Singh}
\affiliation{Institute of Physics, Academia Sinica, Taipei 11529, Taiwan}
\affiliation{Department of Physics, Banaras Hindu University, Varanasi 221005,
India}

\author{Henry T. Wong}
\affiliation{Institute of Physics, Academia Sinica, Taipei 11529, Taiwan}

\author{Chih-Liang Wu}
\affiliation{Department of Physics, National Taiwan University, Taipei 10617, Taiwan}

\author{Chih-Pan Wu}
\affiliation{Department of Physics, National Taiwan University, Taipei 10617, Taiwan}

\begin{abstract}
An \emph{ab initio} calculation of atomic ionization of germanium
(Ge) by neutrinos was carried out in the framework of multiconfiguration
relativistic random phase approximation. The main goal is to provide
a more accurate cross section formula than the conventional one, which is based
on the free electron approximation, for searches of neutrino magnetic
moments with Ge detectors whose threshold is reaching down
to the sub-keV regime. Limits derived with both methods are compared,
using reactor neutrino data taken with low threshold germanium detectors.
\end{abstract}

\maketitle

Neutrino magnetic moments (NMM) describe possible electromagnetic
couplings of the neutrino with the photon via its spin (for reviews, see e.g., Refs.~\cite{Wong:2005pa,Broggini:2012df}).
In the minimally-extended Standard Model (SM), massive neutrinos acquire
non-vanishing, but extremely small, NMMs through electroweak radiative
corrections: $\mu_{\nu}\simeq3\times10^{-19}\,\mu_{\mathrm{B}}\,[m_{\nu}/1\,\mathrm{eV}]$
in units of the the Bohr magneton $\mu_{\mathrm{B}}$
~\cite{Marciano:1977wx,Lee:1977tib,Fujikawa:1980yx}.
The current upper limits set on $\mu_{\nu}$ are orders of magnitude
larger than this SM prediction
. A large NMM,
if observed, will not only imply sources of new physics, but also have significant impact
to the evolution of early Universe and stellar nucleosynthesis 
(see e.g., Ref.~\cite{Fukugita:2003bk}). Furthermore, it might 
favor Majorana neutrinos~\cite{Bell:2005kz,*Bell:2006wi}.

The current experimental limits on $\mu_{\nu}$ are extracted from
the energy spectra of recoil electron in neutrino scattering off detectors.
The scattering cross section contains two incoherent contributions:
one from the weak interaction, $\sigma_{w}$, which preserves the neutrino helicity,
and the other from the magnetic interaction, $\sigma_{\mu}$, which flips it. When the
incident neutrino energy ($E_{\nu}$) and the energy loss to the detector
$(T)$ are high enough so that the binding effects of electrons can
be ignored, the neutrino--free-electron scattering formula is~\cite{Vogel:1989iv}
\begin{align}
\frac{d\sigma_{w}^{(0)}}{dT} & =\dfrac{G_{F}^{2}m_{e}}{2\pi}\left[g_{\nu}^{2}+g_{\nu}^{'2}\left(1-\dfrac{T}{E_{\nu}}\right)^{2}-g_{\nu}g_{\nu}^{'}\dfrac{m_{e}T}{E_{\nu}^{2}}\right],\\
\dfrac{d\sigma_{\mu}^{(0)}}{dT} & =4\pi\alpha\mu_{\nu}^{2}\left(\dfrac{1}{T}-\dfrac{1}{E_{\nu}}\right),
\end{align}
where $G_{F}$ and $\alpha$ are the Fermi and fine structure constants;
the flavor dependent weak couplings, depending on the Weinberg angle
$\theta_{W}$, are $g_{\nu_{e}}=1+2\sin^{2}\theta_{W}$, $g_{\nu_{\mu,\tau}}=-1+2\sin^{2}\theta_{W}$,
$g_{\nu_{e,\mu,\tau}}^{'}=2\,\sin^{2}\theta_{W}$, and interchange
$g_{\bar{\nu}(\nu)}\leftrightarrow g_{\nu(\bar{\nu})}^{'}$ for corresponding
antineutrinos. Based on this formula, several groups recently published
their results: $\mu_{\bar{\nu}_{e}}<2.9\times10^{-11}\,\mu_{\mathrm{B}}$~\cite{Beda:2012zz,*Beda:2013mta}
(GEMMA) and $\mu_{\bar{\nu}_{e}}<7.4\times10^{-11}\,\mu_{\mathrm{B}}$~\cite{Li:2002pn,*Wong:2006nx}
(TEXONO) for reactor antineutrinos, and $\mu_{\nu_{\odot}}<5.4\times10^{-11}\,\mu_{\mathrm{B}}$~\cite{Arpesella:2008mt}
(Borexino) for solar neutrinos. 

One way to improve the experimental sensitivities is to lower the
detector threshold so that events with low $T$ can be registered.
Comparing Eqs.~(1,2), one sees that for $T \ll E_{\nu}$, the weak
part remains constant while the magnetic part increases as $1/T$,
which indicates an enhanced sensitivity to $\mu_{\nu}$. The GEMMA
and TEXONO experiments both used germanium (Ge) semiconductor detectors,
with thresholds at $T=2.8$ and $12\,\mathrm{keV}$, respectively, for
the reported bounds on $\mu_{\bar{\nu}_{e}}$ quoted above. Recently,
the threshold of Ge detectors has been further lowered down to the
sub-keV regime for light WIMP searches and for the studies of neutrino-nucleus
coherent scattering~\cite{Lin:2007ka,Li:2013fla,Zhao:2013xsf}. 

As the kinematics in neutrino scattering with sub-keV energy transfer
starts to overlap with atomic scales, how the atomic binding effects
modify the above free scattering formula becomes an essential issue.
This problem has recently been intensively re-visited because of a
derivation that atomic structure can greatly enhance the magnetic cross
section by orders of magnitude over the free scattering formula at
low $T$~\cite{Wong:2010pb}, in contrast to previous studies all
showing suppression~\cite{Fayans:1992kk,Kopeikin:1997ge,Fayans:2001pg,Kopeikin:2003bx,Gounaris:2001jp}.
While latter works~\cite{Voloshin:2010vm,Kouzakov:2010tx,Kouzakov:2011vx,Chen:2013iud}
justified, with generic arguments and schematic calculations,
that atomic binding effects suppress the scattering cross sections
and the usability of a simple free electron approximation~\cite{Kopeikin:1997ge},
it remains challenging to obtain 
a differential cross section formula at low $T$ with a reasonable
error estimate. In this letter, we address the case of germanium and
report an \textit{ab initio} calculation of germanium ionization by
scattering of reactor antineutrinos 

\[
\bar{\nu}_{e}+\mathrm{Ge}\rightarrow\bar{\nu}_{e}+\mathrm{Ge}^{+}+e^{-}\,.
\]

Taking an ultrarelativistic limit for neutrinos $m_{\nu}\rightarrow0$,
the double differential cross sections for unpolarized scattering
with complex atomic targets are expressed as 
\begin{align}
 & \frac{d\sigma_{w}}{dTd\Omega}\nonumber \\
 & =\frac{G_{F}^{2}}{2\pi^{2}}(E_{\nu}-T)^{2}\cos^{2}\frac{\theta}{2}\bigg[R_{00}^{(w)}-\frac{T}{q}R_{03+30}^{(w)}+\frac{T^{2}}{q^{2}}R_{33}^{(w)}\nonumber \\
 & \quad+(\tan^{2}\frac{\theta}{2}+\frac{Q^{2}}{2q^{2}})R_{11+22}^{(w)}+\tan\frac{\theta}{2}\sqrt{\tan^{2}\frac{\theta}{2}+\frac{Q^{2}}{q^{2}}}R_{12+21}^{(w)}\bigg]\,,\label{eq:sigma_weak}\\
 & \frac{d\sigma_{\mu}}{dTd\Omega}\nonumber \\
 & =\alpha\mu_{\nu}^{2}(1-\frac{T}{E_{\nu}})\bigg[\frac{(2E_{\nu}-T)^{2}Q^{2}}{q^{4}}R_{00}^{(\gamma)}\nonumber \\
 & \quad+\frac{4E_{\nu}(E_{\nu}-T)-Q^{2}}{2q^{2}}\, R_{11+22}^{(\gamma)}\bigg]\,,\label{eq:sigma_mag}
\end{align}
where $\theta$ is the neutrino scattering angle, $q=|\vec{q}|$ is
the magnitude of three-momentum transfer, and $Q^{2}=q^{2}-T^{2}>0$.
The response functions 
\begin{eqnarray}
R_{\mu\nu}^{(w,\gamma)} & = & \frac{1}{2J_{i}+1}\sum_{M_{J_{i}}}\sum_{f}\braket{f|j_{w,\gamma}^{\mu}|i}\sum_{f}\braket{f|j_{w,\gamma}^{\nu}|i}^{*}\nonumber \\
 &  & \times\delta(T+E_{i}-E_{f})\,,
\end{eqnarray}
depending on $q$ and $T$, involve a sum of the final scattering
states $\ket{f}$ and a spin average of the initial states $\ket{i}=\ket{J_{i},M_{J_{i}},\ldots}$,
and the Dirac delta function imposes energy conservation. The relativistic
weak and electromagnetic four-currents are 
\begin{align}
j_{w}^{\mu} & =\bar{e}'[(\frac{1}{2}+2\,\sin^{2}\theta_{W})\gamma^{\mu}-\frac{1}{2}\gamma^{\mu}\gamma_{5}]e\,,\\
j_{\gamma}^{\mu} & =\bar{e}'\gamma^{\mu}e\,,
\end{align}
where the Greek index $\mu=0$ and $1,2,3$ specify the charge and
spatial current densities, respectively, and the direction of $\vec{q}$
is taken to be the quantization axis $\mu=3$. Note that we perform
a Fierz reordering to the weak charged-current interaction (in the
four-fermion contact form) and get a more compact cross section formula in 
Eq.~(\ref{eq:sigma_weak}), in which $j_{w}^{\mu}$ is a sum of the
charged and neutral currents. Also we apply vector current conservation
to relate the longitudinal component $j_{\gamma}^{3}$ to $j_{\gamma}^{0}$,
so the response functions $R_{03,30,33}^{(\gamma)}$ are effectively
included in Eq.~(\ref{eq:sigma_mag}). 

\begin{table*}[t]
\caption{The single-particle energies of Ge atoms calculated by MCDF (s.p.)
versus the edge energies extracted from photoabsorption data (edge)
\cite{Henke:1993gd} of Ge solids. All energies are in units of $\mathrm{eV}$.
\label{tab:sp_energy}}

\begin{ruledtabular}
\begin{tabular}{ccccccccccccc}
 & $K(1s_{\frac{1}{2}})$ & $L_{I}(2s_{\frac{1}{2}})$ & $L_{II}(2p_{\frac{1}{2}})$ & $L_{III}(2p_{\frac{3}{2}})$ & $M_{I}(3s_{\frac{1}{2}})$ & $M_{II}(3p_{\frac{1}{2}})$ & $M_{III}(3p_{\frac{3}{2}})$ & $M_{IV}(3d_{\frac{3}{2}})$ & $M_{V}(3d_{\frac{5}{2}})$ &$N_{I}(4s_{\frac{1}{2}})$ & $N_{II}(4p_{\frac{3}{2}})$ & $N_{III}(4p_{\frac{1}{2}})$\tabularnewline
s.p. & 11185.5 & 1454.4 & 1287.9 & 1255.6 & 201.5 & 144.8 & 140.1 & 43.8 & 43.1 & 15.4 & 8.0 & 7.8\tabularnewline
edge & 11103.1 & 1414.6 & 1248.1 & 1217.0 & 180.1 & 124.9 & 120.8 & 29.9 & 29.3 &  &  & \tabularnewline
\end{tabular}
\end{ruledtabular}
\end{table*}

The many-body theory we adopted in this work to evaluate the germanium
response functions is the multiconfiguration relativistic random-phase
approximation (MCRRPA)~\cite{Huang:1981wj,Huang:1982re}. In essence,
this method is based on the time-dependent Hartree-Fock (HF) approximation, however,
several important features, as the name suggests, make it a better
tool beyond HF to describe transitions of open-shell atoms of high
atomic number $Z$: First, for open-shell atoms, typically there are
more than one configurations which have the desired ground state properties,
therefore, a proper HF reference state should be formed by a linear
combination of these allowed configurations, i.e., a multiconfiguration
reference state. Second, for atoms of high $Z$, the relativistic
corrections can no longer be ignored. By using a Dirac equation, instead
of a Schr{\"o}dinger one, 
the leading relativistic terms
in the atomic Hamiltonian are treated nonperturbatively from the onset.
Third, two-body correlation in addition to HF is generally important
for excited states and transition matrix elements. The random-phase
approximation (RPA) is devised to account for part of the additional
two-body correlation (particles can be in the valence or core states)
not only for the excited but also for the reference state, and in
a lot of cases, it gives good agreement with experiment~\cite{Amusia:1975rev}. 
Furthermore,it has been shown that RPA equations preserve gauge 
invariance~\cite{Lin:1977dl};
this provides a measure of stability of their solutions. 

The MCRRPA has been applied successfully to photoexcitation and photoionization
of divalent atoms such as Be, Mg, Zn, etc.; some of the results are summarized
in~\cite{Huang:1995cc}. 
Following similar treatments, we consider the
electronic configuration of germanium as a core filled up to the $4s$
orbits, with two valence electrons in the $4p$ orbits. As the Ge
ground state is a $^{3}P_{0}$ state, it is a linear combination of
two configurations: $[\mathrm{Zn}]4p_{1/2}^{2}$ and $[\mathrm{Zn}]4p_{3/2}^{2}$.
The wave function is calculated using the multiconfiguration Dirac-Fock
(MCDF) package~\cite{Desclaux:1974jp}. The atomic excitations due
to weak and magnetic scattering are solved by the MCRRPA equation,
and consequently transition matrix elements are yielded. In our calculation,
all the current operators are expanded by spherical multipoles, and
the resulting final scattering states are represented in the spherical
wave basis and subject to the incoming-wave boundary condition. 

Compared with recent work on the same subject~\cite{Fayans:2001pg,Kopeikin:2003bx}
which are also in the similar spirit of relativistic HF, the MCRRPA
approach is refined in several respects: (1) As indicated by the near degeneracy  
of the $N_{II}(4p_{3/2})$ and $N_{III}(4p_{1/2})$ levels in Table~\ref{tab:sp_energy},
using a multiconfiguration reference state is necessary. (2) The non-local
Fock term is treated exactly, without resorting to the local exchange
potentials. (3) The excited states are calculated with two-body correlation
built in by MCRRPA, not simply by solving a Coulomb wave function
with a static one-hole mean field.

To benchmark our Ge calculation, we first list all the single-particle
energies calculated by MCDF and the edge energies extracted from photoabsorption
data~\cite{Henke:1993gd} in Table~\ref{tab:sp_energy}. Although
they are not fully equivalent, good agreements are seen for the inner
shells. The discrepancy in the outer shells mostly comes from the
fact that the data are taken from Ge solids whose crystal structure
is supposed to modify the atomic wave function. As we shall show later,
this is not important for the kinematic range we are interested. On
the other hand, the first ionization energy of the Ge atom in our
calculation $=7.856\,\mathrm{eV}$ agrees with the experimental value
$=7.899\,\mathrm{eV}$~\cite{NIST-ASD}. 

\begin{figure}[h]
\includegraphics[width=1\columnwidth]{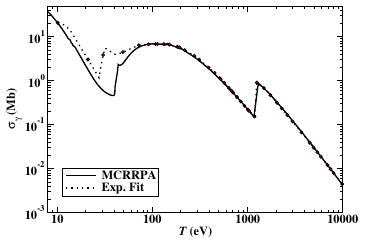}
\caption{Germanium photoionization cross section. The solid line is the result
of our atomic calculation and the dotted curve is the fit of experimental
data (shown in red circles) of Ge solids~\cite{Henke:1993gd}. \label{fig:sigma_gamma}}
\end{figure}

A more definitive test is done with the photoionization process. Unlike
the weak and magnetic scattering by neutrinos where the atom absorbs
a virtual gauge boson, it is a real photon, with $|\vec{q}|=T$, being
absorbed. In Fig.~\ref{fig:sigma_gamma}, the photoionization cross
sections $\sigma_{\gamma}$ for $10\,\mathrm{eV}\le T\le10\,\mathrm{keV}$
from our calculation (for more details, see Ref.~\cite{Long_Paper})
are compared with the fit of experiments~\cite{Henke:1993gd}. Starting
from $T\sim80\,\mathrm{eV}$, our calculation well reproduces the
data curve with an error within $5\%$ in the entire range of $T$
up to $10\,\mathrm{keV}$. For $T<80\,\mathrm{eV}$, the crystal modification
of atomic wave functions becomes important, in particular for the
$3d$ orbit as evidenced by the dislocation of its photoionization
peak. For later calculations of weak and magnetic scattering, we thus
set a minimum of $T_{\mathrm{min}}=100\,\mathrm{eV}$---an already
ambitious threshold for next-generation detectors---so that the atomic
cross section formulae can be applied, and leave the $T<100\,\mathrm{eV}$
region for future study. On the other hand, an important remark is
due here: Photoionization in fact only probes the ``on-shell'' transverse
electromagnetic response functions, i.e., $R_{11+22}^{(\gamma)}|_{q=T}$.
One still needs more experiments to completely check the relevant
response functions, however, this benchmark test does give one confidence
on the applicability of our approach and a realistic error estimate. 

\begin{figure}[h!]
\begin{tabular}{c}
\includegraphics[width=1\columnwidth]{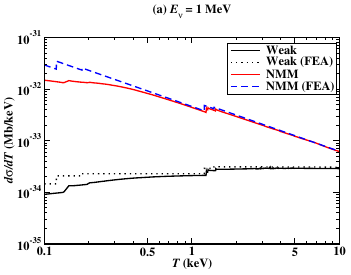}\tabularnewline
\includegraphics[width=1\columnwidth]{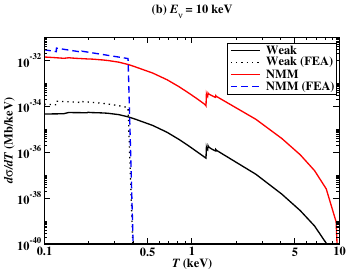}\tabularnewline
\end{tabular}

\caption{The differential cross sections of $\bar{\nu}_{e}$-germanium ionization
with (a) $E_{\nu}=1\,\mathrm{MeV}$ and (b) $E_{\nu}=10\,\mathrm{keV}$.
For magnetic scattering, the neutrino magnetic moment is set to be
the current upper limit $\mu_{\bar{\nu}_{e}}=2.9\times10^{-11}\,\mu_{\mathrm{B}}$.
\label{fig:dsigma}}
\end{figure}

Representative results of our full calculations of $\bar{\nu}_{e}$-germanium
ionization cross sections are shown in Fig.~\ref{fig:dsigma}; the
case with $E_{\nu}=1\,\mathrm{MeV}$ is typical for reactor antineutrinos,
while $E_{\nu}=10\,\mathrm{keV}$ gives an example of low-energy neutrino
sources such as tritium $\beta$ decay ($Q$ value $=18.6\,\mathrm{keV}$),
which is considered as one strong candidate to constrain NMMs~\cite{Giomataris:2003bp,McLaughlin:2003yg}.

\begin{table*}[th!]

\caption{
Limits on NMM at 90\% CL
with selected reactor neutrino data,
comparing  cross-sections derived by 
both MCRRPA and FEA methods.
The projected sensitivities are with the
parameters shown, together with a benchmark
background level of 1/kg-keV-day.
}

\begin{ruledtabular}
\begin{tabular}{lccccc}
Data & Neutrino Flux & Data Strength & Threshold & 
\multicolumn{2}{c}{NMM Limits at 90\% CL ($\mu_{\rm B}$)} \\
& ($\rm{cm^{-2} s^{-1}}$) & (kg-day) & (keV) & FEA & MCRRPA \\ \hline
TEXONO 1kg HPG \cite{Wong:2006nx}  & $6.4 \times 10^{12}$ &  ON/OFF : 570.7/127.8 & 12 
& $< 7.4 \times 10^{-11}$ & $< 7.4 \times 10^{-11}$ \\
TEXONO 900g PPCGe \cite{Li:2013fla} & $6.4 \times 10^{12}$ & ON : 39.5 & 0.5 
& $< 1.6 \times 10^{-10}$ & $< 1.6 \times 10^{-10}$ \\
TEXONO 500g PPCGe & $6.4 \times 10^{12}$ & ON/OFF : 25.5/13.4  & 0.3
& $< 3.0 \times 10^{-10} $ & $< 3.0 \times 10^{-10} $ \\
GEMMA 1.5 kg HPGe \cite{Beda:2012zz} & $2.7 \times 10^{13}$ & ON/OFF : 1133.4/280.4 & 2.8 
& $< 2.9 \times 10^{-11}$ & $< 2.9 \times 10^{-11}$ \\
PPCGe Projected & $6.4 \times 10^{12}$ & (ON/OFF) : 1500/ 500 & 0.3 
& $< 2.3 \times 10^{-11}$ & $< 2.6 \times 10^{-11}$ 
\end{tabular}
\end{ruledtabular}

\label{tab::results}
\end{table*}


As seen from this figure (where $\mu_{\bar{\nu}_{e}}$ is assumed
to be the current upper limit $2.9\times10^{-11}\,\mu_{\mathrm{B}}$),
the sub-keV measurements with Ge detectors can in principle allow
an improved limit by an order of magnitude. On the same plot, we also
compare with the results from the free electron approximation (FEA)~\cite{Kopeikin:1997ge}
\begin{equation}
\frac{d\sigma_{w,\mu}^{(\theta)}}{dT}=\sum_{i=1}^{Z}\frac{d\sigma_{w,\mu}^{(0)}}{dT}\theta(T-B_{i})\,,\label{eq:dsigma_SA}
\end{equation}
in which the free electron formulae, Eqs.(1,2), are used for all electrons
with binding energies $B_{i}$ less than $T$ (implemented by the
theta function). With both $E_{\nu}$ and $T$ bigger than the relevant
atomic scales, it is not a surprise that Eq.(\ref{eq:dsigma_SA})
gives a good description, as illustrated by Fig.2(a) with $T\gtrsim1\,\mathrm{keV}$.
However, as $T$ drops down to the sub-keV regime, the atomic binding
effect starts to manifest and results in suppression of the differential
cross sections, which can be as large as a factor of $0.63$ and $0.5$ for
the weak and magnetic scattering, respectively. On the other hand,
for the case of $E_{\nu}=10\,\mathrm{keV}$, the free electron picture
fails in the entire range of $T$, because the minimum de Broglie
wavelength that could be reached by the incident neutrino $\lambda\sim0.1\,\mathrm{keV}^{-1}$
is not much smaller than the mean orbital radius of Ge $\equiv\sum_{i=1}^{Z}\langle r_{i}\rangle/Z\sim0.2\,\mathrm{keV}^{-1}$.
Furthermore, the free electron dynamics enforces a cutoff for the
maximum of $T_{\mathrm{max}}=2E_{\nu}^{2}/(2E_{\nu}+m_{e})\approx0.38\,\mathrm{keV}$
{[}seen from Fig.2(b){]}, which differs widely from the physical situation.

To compare with experiments, the spectrum-weighed cross section should be used. 
It can be derived from the differential cross-sections of Eqs.~(3,4), giving
\begin{equation}
\left\langle \frac{d\sigma}{dT}\right\rangle =\frac{\int dE_{\nu}\phi(E_{\nu})\frac{d\sigma}{dT}(E_\nu)}{\int dE_{\nu}\phi(E_{\nu})}\,,
\end{equation}
where $\phi(E_{\nu})$ is the neutrino spectrum.

\begin{figure}[h]
\includegraphics[width=8cm]{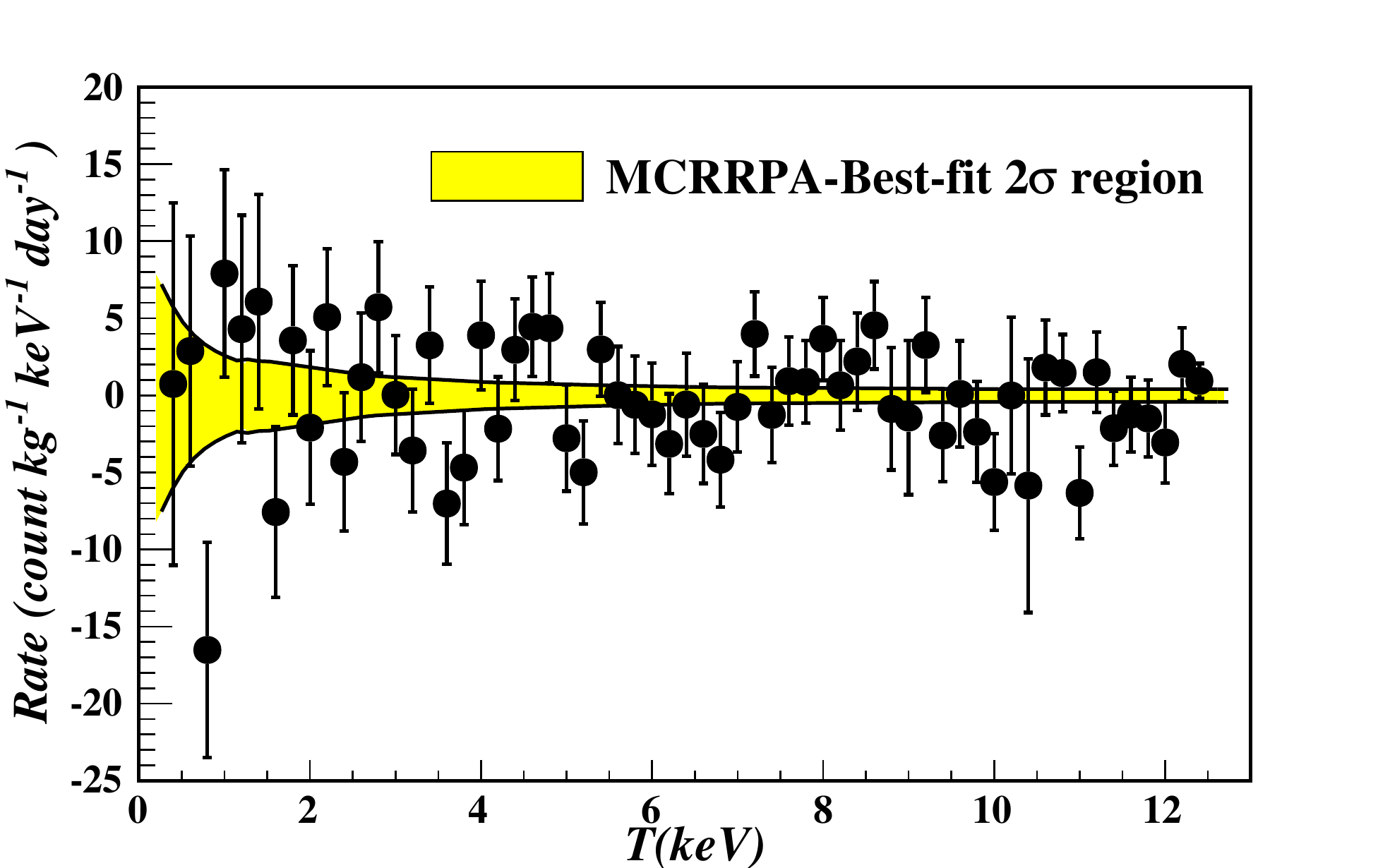}
\caption{
Reactor ON$-$OFF residual spectrum with a
PPCGe from 25.5/13.4 kg-day of ON/OFF data at KSNL
at a threshold of 300~eV. The two-sigma
allowed band from MCRRPA anlaysis is
also displayed
}
\label{fig::texono13}
\end{figure}

Analysis was performed with data taken with standard high-purity ermanium (HPGe)
and $p$-type point-contact germanium detectors (PPCGe) with sub-keV sensitivity 
at the Kuo-Sheng Reactor Neutrino Laboratory (KSNL)~\cite{Wong:2006nx,Li:2013fla}. 
The key experimental parameters and the 90\% CL limits are summarized in 
Table~\ref{tab::results}, 
for both MCRRPA and FEA methods. Also listed are the published FEA and derived MCRRPA bounds by the
GEMMA experiment~\cite{Beda:2013mta}, and 
the projected sensitivities for PPCGe under realistic conditions. 
The TEXONO PPCGe Reactor ON$-$OFF spectrum  with PPCGe 
from 25.5/13.4 kg-day of ON/OFF data at a threshold of 300~eV
and the corresponding NMM squared constraints
are displayed in Fig.~\ref{fig::texono13}.

In summary, we demonstrate in this work that by using the multiconfiguration relativistic
random phase approximation, the atomic structure of germanium and its photoabsorption data
with photon energy larger than 100~eV can be reliably calculated. Applying the method to
the atomic ionization by the neutrino weak and magnetic moment interactions, it is found 
that while the conventional scattering formula based on the free electron approximation 
works reasonably well when the neutrino energy loss is larger than 1~keV, the atomic effect 
starts to play a significant role for sub-keV energy loss. With new-generation germanium 
detectors lowering their thresholds down to the sub-keV regime and enhancing their 
sensitivities to neutrino magnetic moments, our scattering formulae should provide more 
reliable constraints.

\begin{acknowledgments}
We acknowledge the supports from the National Science Council, Republic Of China under Grant Nos. 
102-2112-M-002-013-MY3 (JWC, CLW, CPW), 98-2112-M-259-004-MY3 (CPL), and 101- 2112-M-259-001 (CPL); 
the CTS and CASTS of NTU (JWC, CLW, CPW). 

\end{acknowledgments}

\bibliographystyle{apsrev4-1}
\bibliography{Ge_NMM}

\end{document}